\documentclass[aps,prb,twocolumn]{revtex4-1}
\usepackage[utf8]{inputenc}
\usepackage{xcolor}
\usepackage{amsmath,amssymb}
\usepackage{graphicx}

\begin{document}

% Use the \preprint command to place your local institutional report
% number in the upper righthand corner of the title page in preprint mode.
% Multiple \preprint commands are allowed.
% Use the 'preprintnumbers' class option to override journal defaults
% to display numbers if necessary
%\preprint{}

%Title of paper
\title{Quantum selection for spin systems}

% repeat the \author .. \affiliation  etc. as needed
% \email, \thanks, \homepage, \altaffiliation all apply to the current
% author. Explanatory text should go in the []'s, actual e-mail
% address or url should go in the {}'s for \email and \homepage.
% Please use the appropriate macro foreach each type of information

% \affiliation command applies to all authors since the last
% \affiliation command. The \affiliation command should follow the
% other information
% \affiliation can be followed by \email, \homepage, \thanks as well.
\author{Alix Deleporte}
%\email[]{Your e-mail address}
%\homepage[]{Your web page}
%\thanks{}
%\altaffiliation{}
\affiliation{Université de Strasbourg, CNRS, IRMA UMR 7501, F-67000 Strasbourg,
France}

%Collaboration name if desired (requires use of superscriptaddress
%option in \documentclass). \noaffiliation is required (may also be
%used with the \author command).
%\collaboration can be followed by \email, \homepage, \thanks as well.
%\collaboration{}
%\noaffiliation

\newcommand{\R}{\mathbb{R}}
\renewcommand{\S}{\mathbb{S}}
\newcommand{\todo}[1]{\textcolor{orange}{TODO:~ #1}~}
\newcommand{\citer}{\todo{Citer}}
\newcommand{\refe}{\todo{Ref}}

\date{\today}

\begin{abstract}
We report mathematical results on the process by which quantum order by disorder
takes place for spin systems. The selection rules follow the influence
of several competing contributions. Moreover there is no link between
quantum selection and thermal selection. We present work in the
general setting as well as toy models and examples for which quantum
selection has an interesting behaviour.
\end{abstract}

% insert suggested PACS numbers in braces on next line
% insert suggested keywords - APS authors don't need to do this
%\keywords{}

%\maketitle must follow title, authors, abstract, \pacs, and \keywords
\maketitle

% body of paper here - Use proper section commands
% References should be done using the \cite, \ref, and \label commands
\section{Introduction}

\subsection{Order by disorder}
\label{sec:order-disorder}

The understanding of low-energy states of non-integrable quantum systems is a
notoriously difficult task, with applications to the design of both
quantum and regular computers, supraconductivity as well as
superfluidity. In particular, the Anderson RVB model for high $T_c$
supraconductivity \cite{anderson_resonating_1973} has drawn attention to frustrated quantum spin
systems.

In an effort to tackle this problem from a theoretical perspective,
various approximation procedure are used, such as restriction to
finite size systems \cite{lecheminant_order_1997,waldtmann_first_1998,depenbrock_nature_2012} or generalizations to $SU(N)$ with $N$ large \cite{sachdev_kagome-_1992}. In this paper
we are interested in semiclassical methods\cite{harris_possible_1992,chubukov_order_1992,doucot_semiclassical_1998}, which are inspired by
Villain's ``order by disorder'' principle\cite{villain_order_1980}.

\begin{figure}
  \centering
  \includegraphics[width=0.4\linewidth]{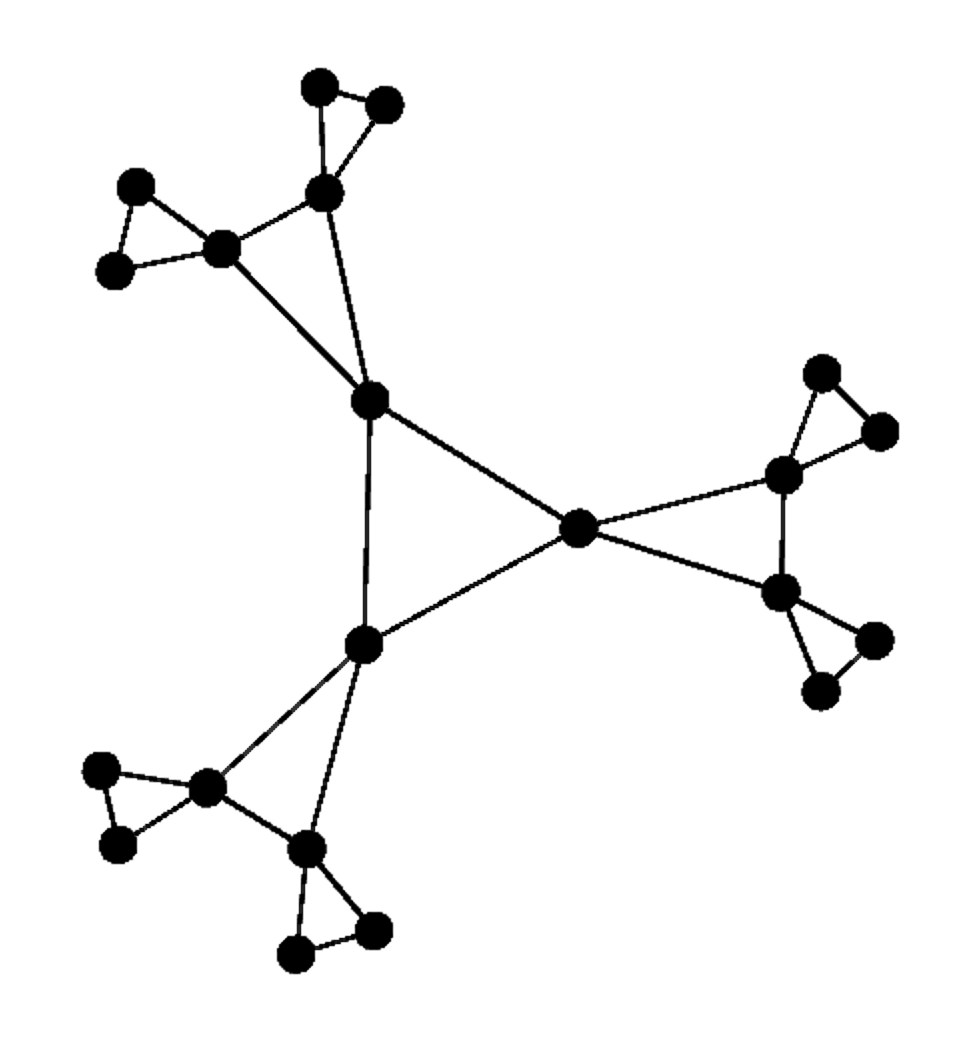}
  \includegraphics[width=0.5\linewidth]{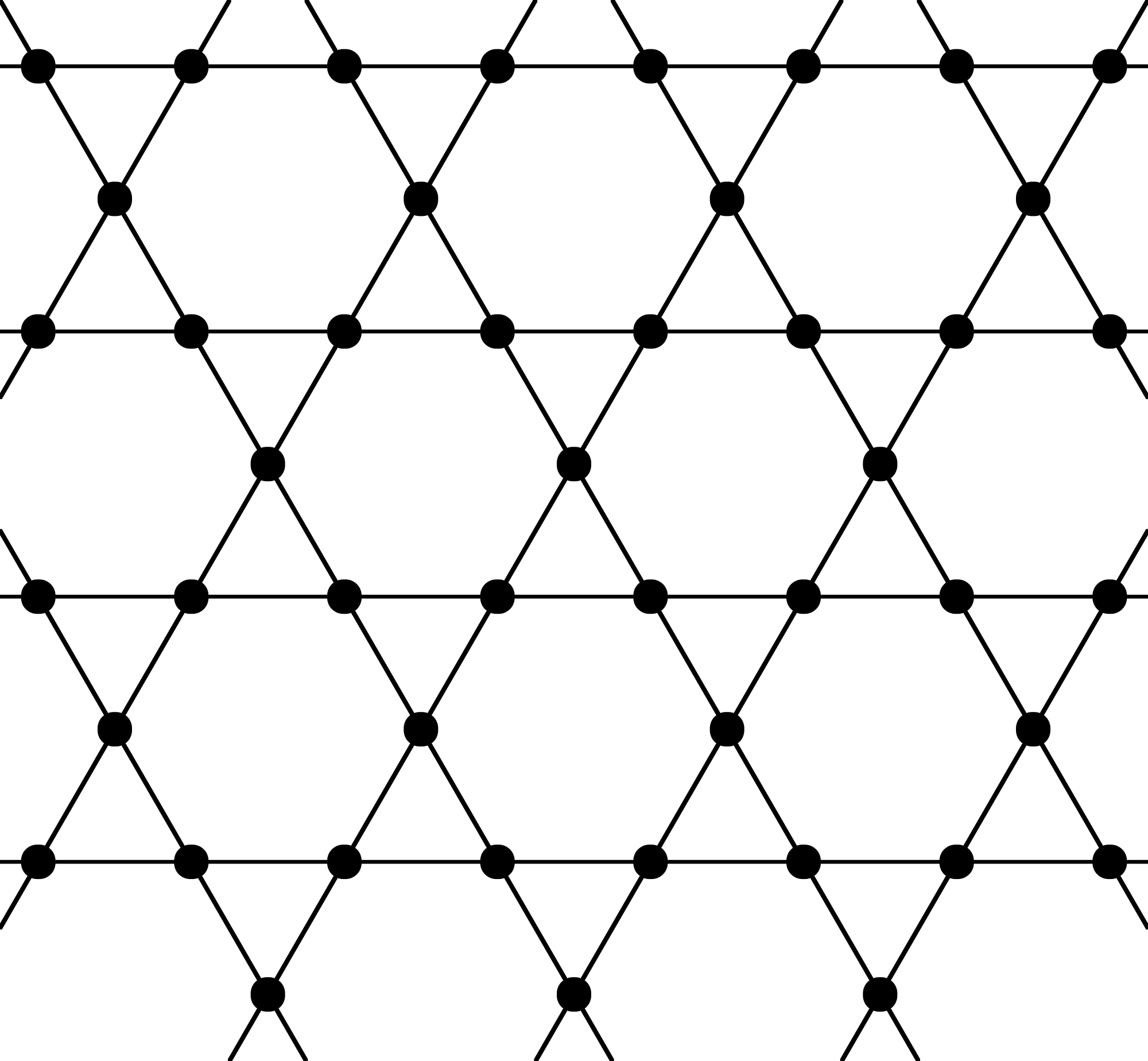}
  \caption{Pieces of the Husimi tree (left) and the Kagome lattice
    (right)}
  \label{fig:huskag}
\end{figure}

This approach is motivated by the fact that, for frustrated spin
systems, the classical minimal set does not consist of a single
class of configurations given by a global symmetry. Spin ices
\cite{anderson_ordering_1956,matsuhira_low-temperature_2002,harris_geometrical_1997,lago_cder2se4_2010} feature a discrete set of classical minimal configurations, with
extensive cardinality. For the Heisenberg AntiFerromagnetic model (HAF) on
the Kagome lattice (see Figure \ref{fig:huskag}), they form a continuous
set which is not regular: the dimension of allowed infinitesimal moves
is not constant on this set.

The idea behind ``order by disorder'' is that low-temperature
classical states, as well as quantum low-energy eigenstates, are not exactly
located on the classical minimal set but are spread out; in
particular, their energies are shifted up by a
factor depending on the behaviour of the classical energy
near its minimal set. The flattest the classical energy landscape, the
lowest the energy contribution. As a consequence, those states must
concentrate only on the subset of the classical minimal set where the
local energy landscape is the flattest. In short, the presence of
thermal or
quantum fluctuations actually restrict the possible locations of
low-energy states.

At this point we already make an emphasis on the geometrical data
needed to define what it means for the classical energy to be flatter
near one minimal point than near another. As was already pointed out\cite{doucot_semiclassical_1998},
in the setting of classical
low-temperature, the Gibbs measure depends on the classical energy
itself and the volume element on the phase space. To the contrary,
quantum states depend on the symplectic structure on the phase space,
which is a finer geometrical notion: some phase space transformations
preserve the volume form but not the symplectic structure. Thus,
though thermal and quantum selection stem from the same intuition, the
``flattest'' classical points may not be the same in the two cases.

% \subsection{Frustrated spin systems}
% \label{sec:frustr-spin-syst}
% Frustrated spin systems are antiferromagnetic spin systems on lattices
% which do not allow bipartite orderings of the spins. The standard
% example is the triangular lattice, on which the Ising problem admits
% an exact solution. An example of great theoretical and experimental
% interest is the Kagome lattice, which is a tiling of
% the plane with regular hexagons and equilateral triangles
% (see Figure \ref{fig:huskag}). The classical ground state for the Ising or
% Heisenberg antiferromagnet (HAF) is non-trivial on this lattice, since each
% pair of neighboring spins cannot be opposite.

% The set of classical minimal energy of the HAF on the Kagome lattice is a degenerate
% set. Indeed, the energy is minimal when, on each triangle of bonds,
% the spins (which are elements of the two-sphere) form a great
% equilateral triangle on the sphere; this set is highly degenerate as,
% for each pair of connected triangles, there is a degree of freedom in
% the choice of the spins.
% In the case of the Kagome lattice, the phase space set of
% minimal classical energy is not even a smooth manifold: it is believed
% that it is a stratified manifold, as for instance the boundary of a
% hypercube.

% This high classical degeneracy is believed to be reflected in the
% quantum picture. It has been reported \cite{wills_structure_1996}
% that the spin $\frac 12$ HAF has a glassy behaviour at low
% temperature, with a large number of metastable states. The properties
% of the ground state for this model are unknown.

\subsection{Results}
\label{sec:results}
In this article we clarify the process under which quantum selection
takes place, and examine the links with thermal selection. A common
heuristics states that quantum selection and thermal selection follow
the same rules. This intuition, which leads to the claim that on the
Kagome HAF low-energy states are coplanar, is sometimes misleading. Another
claim states that quantum selection is determined by the classical
frequencies in the linear spin-wave approximation. In fact there are
additional terms, which do not play a role on antiferromagnetic systems
but which appear in more general spin systems. We describe in
detail those additional terms.

We report mathematical results, which define a function $\mu $ under
which quantum selection takes place: as the spin grows, low-energy
quantum states localize on the set of phase space on which both the
classical energy and this function $\mu $ are minimal. We then analyse
various model situations of irregular minimal classical sets in order
to understand the link with thermal selection.

This article is organised as follows: Section II presents the general
mathematical framework for the treatment of quantum selection in the
context of spin systems. In Section III we use three toy models
to illustrate the concepts and difficulties associated with quantum
order by disorder. In Section IV we analyse practical examples such as
the semiclassical HAF on the Kagome lattice. Section V presents a
discussion of the consequences and applications of our work. The
Appendix consists of exact computations which relate spin systems and
Toeplitz quantization.

%%% Local Variables:
%%% mode: latex
%%% TeX-master: "Main"
%%% End:

\section{Spin wave frequencies and Toeplitz quantization}
In this section we expose the main mathematical ideas behind Toeplitz
quantization, which allows to study spin operators in the large $S$
limit from a rigorous point of view. We report our recent results on
the topic and clarify the exact procedure under which quantum
selection takes place.

\subsection{Harmonic oscillators in Bargmann-Fock representation}
\label{sec:harm-oscill-bargm}
The point of view of Bargmann on the quantum harmonic
oscillator\cite{bargmann_hilbert_1961,husimi_formal_1940}, is
that quantum states should be seen as holomorphic functions on the
complex space $\mathbb{C}^d$ instead of the common choice
$L^2(\mathbb{R}^d)$. This idea can in fact be generalized to other
phases spaces than $\mathbb{C}^d$, and allows to understand the large
spin limit as a semiclassical limit from a rigorous point of view.

For positive $k$ (which is seen as the inverse Planck constant), holomorphic functions on $\mathbb{C}^d$ form a
Hilbert space $B_k$ with the following
scalar product
\[
  \langle u|v\rangle_{B_k}=\int
  \overline{u}(z)v(z)\exp(-k{|z|^2})dz.
\]
We naturally exclude from the space $B_k$ the functions with infinite
norm. Example of functions in $B_k$ are the monomials $z\mapsto
z_1^{\nu_1}\ldots z_d^{\nu_d}$ which, once normalized, form a Hilbert
base of $B_k$.
Under this definition, $B_k$ naturally sits inside the space $L_k$
of all (not necessarily holomorphic) functions which are
square-integrable with respect to the exponential weight above. The
orthogonal projector $\Pi_k$ from $L_k$ to $B_k$ is used to define the
quantum harmonic oscillator, which is the following operator on $B_k$:
\[
  T_k(|z|^2)|u\rangle=\Pi_k(z\mapsto |z|^2u(z)).
\]
This is a \emph{Toeplitz} operator: the composition of a
multiplication operator and a projection. The matrix elements are
simply
\[
  \langle
  u|T_k(|z|^2)|v\rangle_{B_k}=\int_{\mathbb{C}^d}\overline{u}(z)v(z)|z|^2\exp(-k|z|^2)dz.
\]
The monomials $z^{\nu}$ are eigenfunctions of this operator, with
eigenvalues $k^{-1}(\nu_1+\ldots+\nu_d+d)$. This contrasts with the
$L^2(\mathbb{R}^d)$ point of view on the harmonic oscillator, where
the eigenvalues are the half-integers $\hbar(\nu_1+\ldots+\nu_d+\frac
d2)$. This does not mean that this Toeplitz operator is not natural,
or that other terms should be added; in experiments one can only
measure gaps between eigenvalues, which coincide for the two settings.

The definition of the Toeplitz operator can be generalized. If $H$ is
any function on $\mathbb{C}^d$ (which represents the classical energy
on the phase space $\mathbb{C}^d=\mathbb{R}^{2d}$), the associated
Toeplitz operator on $B_k$ is defined as
\[
  T_k(H)|u\rangle=\Pi_k(z\mapsto H(z)u(z)).
\]
This defines a quantization: $T_k(H_1+H_2)=T_k(H_1)+T_k(H_2)$, and in the large $k$ limit, the commutator
$[T_k(H_1),T_k(H_2)]$ becomes close to $-ik^{-1}T_k(\{H_1,H_2\})$. The
function $H$ associated with the operator $T_k(H)$, which is unique,
is called the \emph{symbol} of $T_k(H)$. In the context of spin
systems it coincides with the notion of upper symbol.

Toeplitz quantization follows the Wick order: if $H:z\mapsto
\overline{z}^{\alpha}z^{\beta}$, then \[T_k(H)=k^{-\sum
    \alpha_i}\partial^{\alpha}z^{\beta}.\] The Wick rule allows
explicit computations for the Toeplitz quantization of any polynomial
function in the coordinates.

Of great interest are Toeplitz operators associated with semipositive
definite forms $Q\geq 0$. As in the harmonic case, the infimum of the
spectrum is linked with the classical frequencies, but is shifted with
respect to the usual quantization procedure: if $\lambda_1,\ldots,
\lambda_r$ are the non-zero classical frequencies for $Q$, then
\begin{equation}\label{eq:mu}
\mu(Q):=\inf Spec(T_k(Q))= k^{-1}\hspace{-0.4em}\left(\frac 12\sum_{i=1}^r\lambda_i+\frac 14
  tr(Q)\hspace{-0.2em}\right)\hspace{-0.2em}.
\end{equation}

The factor $tr(Q)$ is specific to Bargmann quantization. In the Weyl
representation, one has instead
\[
  \inf Spec(Op_W^{\hbar}(Q))=\frac{\hbar}{2}\sum_{i=1}^r\lambda_i.
  \]

\subsection{Toeplitz operators on spheres}
\label{sec:toepl-oper-sphere}
Toeplitz quantization can be generalized from $\mathbb{C}^d$ to other phase
spaces, using tools of complex geometry\cite{charles_berezin-toeplitz_2003}. In particular, this allows to
define a quantization procedure on product of spheres: to any
classical energy on a product of spheres, and any $k$, one can
associate a quantum operator, acting on the tensor product of spaces
$\mathbb{C}^{k+1}$. Previously $k$ was any positive real number, but now it needs to be
an integer: the topology of the phase space only allows quantized values of the
inverse Planck constant.

Toeplitz operators on product of spheres include spin systems (with
spin $S=\frac k2$). However
the quantization procedure requires some care in the computations as can be seen on Table \ref{tab:sphere}.
\begin{table}[h]
\begin{center}
  \begin{tabular}{|c|c|}
    \hline
    classical & quantum ($S=\frac k2$)\\
    \hline
    $z$ & $\frac{k}{k+2}S_z$\\
%    $z+2k^{-1}z$ & $S_z$\\
    $x$ & $\frac{k}{k+2}S_x$\\
%    $x+2k^{-1}x$ & $S_x$\\
    $z^2$ & $\frac{k^2}{(k,3)}S_z^2+\frac{1}{k+3}$\\
%    $z^2+k^{-1}(5z^2+1)+k^{-2}(6z^2+2) $ & $S_z^2$\\
    $zx$ & $\frac{k^2}{2(k,3)}(S_xS_z+S_zS_x)$\\
    $z^2x$ & $\frac{k^3}{(k,4)}S_zS_xS_z+\frac{1}{(k+3)}S_x$\\
    $z^3$ & $\frac{k^3}{(k,4)}S_z^3+\frac{k(3k+8)}{(k,4)}S_z$\\
\hline
\end{tabular}
  \caption{  \label{tab:sphere}Quantization of some symbols on the sphere.\\The
    operator $S_z$ has entries
    $-1,-1+S^{-1},\ldots,1-S^{-1},1$.\\We denote
    $(k,j)=(k+2)(k+3)\ldots (k+j)$.}
  \end{center}

\end{table}

The corrective terms of order $k^{-1}$ are crucial for quantum order from disorder.
The details for the computations in Table \ref{tab:sphere} are presented
in the Appendix.

\subsection{Quantum selection for Toeplitz operators}
\label{sec:quant-select-toepl}

In a recent paper\cite{deleporte_low-energy_2016}, we developed mathematical tools in order to study
quantum selection for general Toeplitz operators in the large
$k$ limit.
We report that, in a general case (even if the set of minimal
classical energy is irregular), quantum selection takes place for
Toeplitz operators following a general criterion.

In order to apply our results to usual spin operators, as seen above, we need
to consider Toeplitz operators with classical energy depending on $k$ in the
following way:
\[
  f=f_0+k^{-1}f_1+k^{-2}f_2+\ldots,
\]
where each term $f_j$ is a real function on the phase space. Indeed,
the quantization of symbols which do not depend on $k$ only yield a
deformation of the usual spin operators. The
Toeplitz operator $T_k(f)$ is well-defined by linearity.

Quantum states with energy less than $\min(f_0)+Ck^{-1}$ are known to
localize on $Z=\{f_0\text{ is minimal}\}$ as $k$ grows. In a neighbourhood of any
point $P_0$ of $Z$, the function $f_0$ can be approximated by its
quadratic Taylor estimate $\min(f_0)+Q$, where $Q$ is a semidefinite
positive quadratic form which depends on $P_0$.

The selection criterion is then \[\tilde{\mu}=\mu(Q)+f_1,\] in
following sense: if $(u_k)$ denotes a sequence of ground states of
$T_k(f)$, if a set $V$ lies at positive distance from \[\{x\in Z,
  \tilde{\mu}(x)=\min(\tilde{\mu})\},\] then for
every $j$ one has, as $k\to +\infty$,
\[
  \int_V |u_k(z)|^2 \precsim k^{-j}.
\]
The meaning of $|u_k(z)|^2$ depends on the underlying manifold (for
instance, on $\mathbb{C}^d$ a factor $\exp(-k|z|^2)$ must be added), but as
our quantum states are defined on the whole phase space, localisation
properties can be formulated in a more elementary way than in the
space representation.

The
quantum ground state localizes, in the large $k$ limit, only on the
part of $Z$ where $\tilde{\mu}$ is minimal; at any positive distance
from this set, the ground state decays faster than any negative power
of $k$. In fact, if $E_0$ is the
energy of the ground state, then any quantum eigenstate with energy
less than $E_0+\epsilon k^{-1}$ for any $\epsilon $ localizes where
$\tilde{\mu}$ is minimal.\footnote{If the minimal set $Z$ is infinite,
  the number of eigenstates with energy less than $E_0+\epsilon
  k^{-1}$ for any $\epsilon$ tends to $+\infty$ as $k\to \infty$.}

In order to apply this result from a standard ``operator-presented''
quantum spin Hamiltonian in the large spin limit, one needs first to
compute, not only the associated classical energy at the main order,
but also the so-called ``subprincipal symbol'' which contains the
next-order terms in the quantization procedure. For instance, starting
with the operator $S_z^2$, the principal symbol is of course $z^2$,
and from Table \ref{tab:sphere} one can compute that a more accurate
representation is \[z^2+k^{-1}(5z^2+1).\] This subprincipal part,
added to the trace and to the sum of symplectic eigenvalues of the
quadratic part of the energy, yields the function $\widetilde{\mu}$
which is the selection rule. In section \ref{sec:toy-models} we apply
this method to several models.

The physical interpretation of $\widetilde{\mu}$ is the
following: suppose that one wants to minimise the energy of a quantum
state while constraining it to be localised at a precise point, where
the classical energy has a local minimum. Then the energy of this
minimal constrained state is naturally close to the classical energy,
but is lifted up by quantum fluctuations. Indeed, quantum states have
to spread out somewhat, and to reach parts of the phase space where the
classical energy is not minimal. This energy lift is of the same order
as the semiclassical parameter (here, $k^{-1}$). Then
$\widetilde{\mu}$, at this point, is the $k^{-1}$ contribution to this
energy lift.

In the context of spin systems, the selection rule is determined by
the classical frequencies of the spin waves, and by non-trivial
additional terms which must be taken care of. For the particular case
of HAF systems, if each spin has the same number of neighbors, then
the additional terms are constant, but on other systems on which
quantum selection is studied, they can play an
important role.

%%% Local Variables:
%%% mode: latex
%%% TeX-master: "Main"
%%% End:

\section{Toy models}
\label{sec:toy-models}
In order to understand quantum selection in the general case and in
the particular case of spin systems, we first look at three simple toy
models.

In the first toy model, which is the first historical example of
quantum selection, thermal and quantum selection play the same
role. In the second toy model, which has an irregular minimal set as
does the HAF on the Kagome lattice, thermal selection is sharper than
quantum selection: some classical configurations are equivalent from a
quantum point of view (they share the same value of $\widetilde{\mu}$), but are discriminated by the Gibbs
measure. Conversely, on the third toy model, there is no thermal
selection, but quantum selection takes place.

\subsection{Miniwells}
\label{sec:miniwells}
The general study of quantum selection for the ground state of a
Schr\"odinger operator was performed by Helffer and Sj\"ostrand \cite{helffer_puits_1986}
who exposed a WKB construction for a quasimode associated to the
lowest energy. Quantum selection occurs when the potential $V$ is
minimal on a degenerate set $Z$. If $Z$ is a smooth manifold on which
$V$ vanishes at order 2, the criterion for quantum selection is the
trace of the square root of the Hessian matrix of $V$ at the minimal
points; in this Weyl setting, it corresponds
exactly to the sum of the classical frequencies for the linearized
system. Even in this case it does not correspond to the criterion for
thermal selection (which is the product of these frequencies).

The simplest example is the operator $P_{\hbar}=-\hbar^2\Delta+V$
acting on $L^2(\mathbb{R}^2)$, with $V(x,y)=y^2(1+x^2)$, vanishing at order
two on the horizontal axis. It is already interesting to note that,
though $V$ itself is not a confining potential, $P_h$ only has
discrete spectrum because of the quantum selection.

Around every point $(x_0,0)$ of the horizontal axis, the quadratic
terms in the potential are $y^2(1+x_0^2)$. For this quadratic
potential there is one non-zero classical frequency,
$\sqrt{1+x_0^2}$. This frequency is minimal at $x_0=0$, which is
called the ``miniwell'' for this potential. Hence, the ground
state of this operator concentrates on the point $(0,0)$, in the
previous sense (for the Husimi transform).

In this setting, the value $\widetilde{\mu}$ coincides with the
effective potential given by the intuition of the adiabatic approximation
\cite{doucot_semiclassical_1998}. At $(x_0,0)$ one can approximate the
behaviour of a low-energy state in the second variable as the ground
state of the quadratic transverse operator
$-\hbar^2\partial_y^2+(1+x_0^2)y^2$; if $e_{x_0}(y)$ is the ground
state of this operator, then the energy of a state of the form
$e_{x}(y)f(x)$ is \[\hbar\langle
  f,(-\hbar\partial_x^2+\widetilde{\mu})f\rangle,\] so that
$\widetilde{\mu}$ acts as an effective potential (with new
semiclassical parameter $\sqrt{\hbar}$).
\subsection{Crossing points}
\label{sec:crossing-point}
With the Kagome lattice in mind, let us consider toy models where
the minimal set of the classical energy is not a smooth manifold.

The first of this model is again a Schr\"odinger operator on $\mathbb{R}^2$,
with potential $V(x,y)=x^2y^2$. The minimal set consists in the two
axes, which meet at zero. On the horizontal branch $(x_0,0)$, there is only one
non-zero linear classical frequency, which is $|x_0|$. This frequency
is minimal at zero (note that this frequency is not smooth at
zero). The same applies for the vertical axis. Once again, the
operator $P_{\hbar}=-\hbar^2\Delta+V$ only has discrete spectrum and
the first eigenfunction localizes at the origin, which is also the
point of thermal selection (since the local dimension of the zero
modes is maximal at this point). Because of the non-regularity of the
classical frequency at the crossing point, for a potential $W$ close to
$V$ which is also minimal on the two axes, the quantum system will
still select the crossing point.

In this setting, the Born-Oppenheimer approximation fails at the
crossing point, so that there is no simpler effective
model. However $\widetilde{\mu}$ still acts as an energy barrier,
independently on the geometry.

A more general crossing is the Schr\"odinger operator on $\mathbb{R}^3$, with
potential $V(x,y,z)=x^2y^2z^2$. The minimal set is the union of the
three planes $\{x=0\},\{y=0\},\{z=0\}$, and the local zero dimension
is maximal at the origin. However, on the plane $\{x=0\}$ the
classical frequency is $|yz|$. All classical frequencies vanish identically on the three
axes.

For this particular potential, there is a hierarchy of
perturbations, and investigating the sub-sub-principal (order
$k^{-2}$) terms will lead to concentration at the origin and discrete spectrum. However, if
non-degenerate transverse modes are added, they correspond (in the
adiabatic regime) to a perturbation of order $k^{-1}$, in front of
which the $k^{-2}$ confinement at the origin is negligible; in the
general setting, even for small perturbations, the quantum selected point can be any point on the
three axes. This illustrates the
discrepancy between quantum and thermal selection and shows that for
the Kagome lattice, the points of quantum selection might not
necessarily be the planar configurations, though those configurations
have the maximal number of zero modes.

\subsection{Cancelling terms}
\label{sec:cancelling}

We propose an example which serves to illustrate the effects of
the different terms in the process by which quantum selection takes place.

Let us consider the following one-spin Hamiltonian:
\[
  H=S_z^2+\Delta S_zS_xS_z.
\]
The principal symbol of this operator is
\[
  h_0=z^2(1+\Delta x).
\]
If $0<\Delta<1$, then $h_0$ is minimal on $\{z=0\}$. It looks like
$h_0$ is smaller near $(-1,0,0)$ than near any other point, so that,
at first sight, quantum order from disorder seems to take place in
this setting.

Let us look at the three terms appearing in quantum selection:
\begin{enumerate}
\item For any minimal point, the associated linear classical frequency
  is zero, since the linear classical approximation is that of a free massive
  particle in one-dimensional space.
\item The trace of the quadratic form near a minimal point is
  non-zero; this term has contribution
  \[\mu=1+\Delta x
  \]
\item The next-order term in the expansion is
  \[
    h_1=5z^2-1+\Delta(9z^2x-x)
  \]
  In particular, on the set where $h_0$ is minimal, one has
  \[
    h_1=-1-\Delta x.
  \]
\end{enumerate}
The set selected by quantum order by disorder is the set where
$h_1+\mu $ is minimal, but the two terms cancel out. Hence there is no
quantum selection at this order of expansion.

It can readily be seen that, if the spin $S$ is even, then
$S_z|0\rangle=0$, so that $H|0\rangle=0$ and $|0\rangle $ is the
ground state of $H$. As expected, the magnetization of $|0\rangle $
along the $x$ axis is exactly zero. Hence, there is no quantum
selection in the large spin limit for this model.

More involved theoretical examples where the classical degeneracy is
not lifted at any order of $S^{-1}$ include spin
textures\cite{doucot_large_2016}. In other situations, there could be
no quantum selection at first order, but next-order terms could break
the degeneracy. In practice, one expects additional terms (such as
second nearest neighbours interactions) which will destroy exact degeneracies.

%%% Local Variables:
%%% mode: latex
%%% TeX-master: "Main"
%%% End:

\section{Examples}
\label{sec:examples}

\subsection{Kagome lattice}
\label{sec:kagome-lattice}
The quantum HAF on the Kagome lattice is the
Toeplitz quantization of the classical energy
\[
  \sum_{i\sim j}\mathbf{e}_i\cdot \mathbf{e}_j.
\]

Here $i\sim j$ means that the two sites $i$ and $j$ are linked by an
edge. The Toeplitz quantization of the symbol above is
\[
  \cfrac{S^2}{(S+1)^2}\sum_{i\sim j}S_i\cdot S_j,
\]

so that, up to a multiplicative factor, the quantization of the
classical HAF is the quantum HAF. As we wish to study quantum
selection, it is important that in this case $h_1=0$.

The low-temperature properties of the $S=\frac 12$ HAF are still
unknown. Various numerical Ans\"atze or exact diagonalizations\cite{iqbal_spin-1_2015,iqbal_vanishing_2014,iqbal_gapless_2013,zeng_quantum_1995,waldtmann_first_1998,lecheminant_order_1997,depenbrock_nature_2012}
predict a spin liquid phase with polynomial decay of correlations,
which is consistent with experiments\cite{wills_structure_1996}. The
large $S$ analysis sheds some light on this behaviour as we will see.

If $\mathbf{e}_i=(x_i,y_i,z_i)$, since sites in the Kagome lattice are
connected in triangles, up to a constant the classical energy reads
\[
  \sum_{\text{triangles}}\|\mathbf{e}_i+\mathbf{e}_j+\mathbf{e}_k\|^2.
\]
The minimal classical set consists in configurations where, on each
triangle of sites, spins form a great equilateral triangle on the
sphere. This set has a highly non-trivial structure: the presence of
loops of triangles makes it non-smooth. The classical minimal set for
one hexagon of triangles already has a crossing point as one of the
toy models, on which two smooth manifolds cross.

An interesting subset of classical minimal configurations consists in
planar configurations, which form a discrete set. It is believed that
quantum order by disorder selects these configurations, thus reducing
the semiclassical study to a $3$ colours Potts model on the Kagome
lattice (with Hamiltonian unknown so far). Some of those planar
configurations have been proven \cite{chubukov_order_1992} to be local minima for the function
$\mu$ which is the criterion for quantum selection, but it is unknown
whether these are the global minima for $\mu$ or not. The results in
the $S=\frac 12$ case are compatible with this approach as there is an
extensive number of coplanar states, the majority of which having no
long-range order; this contrasts with the $SU(N)$-case
\cite{sachdev_kagome-_1992} which would predict a unique, ordered,
selected configuration.

\subsection{Simple models for the Kagome lattice}
\begin{figure}
  \centering
  \includegraphics[width=0.45\linewidth]{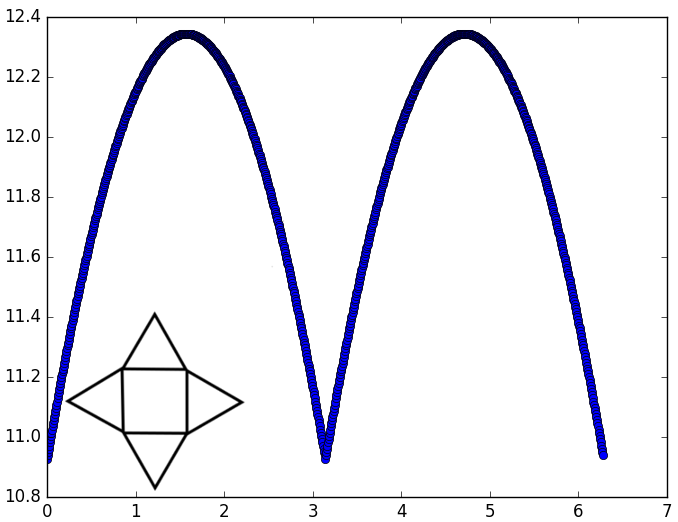}
  \includegraphics[width=0.45\linewidth]{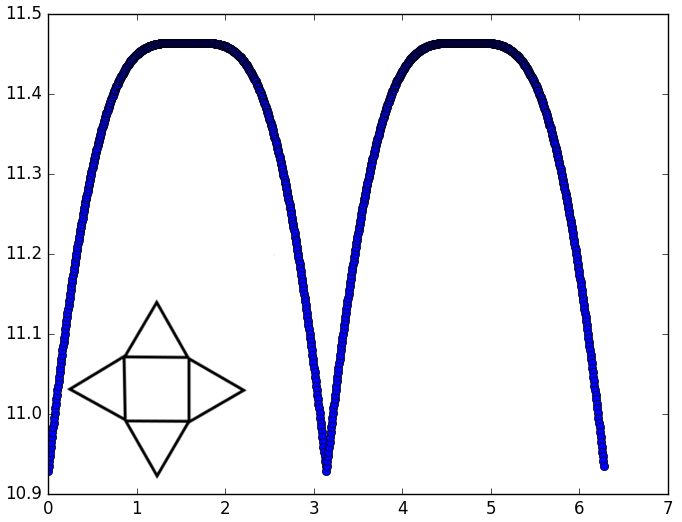}
  \caption{Numerical plot of $\mu$ for a loop of 4 triangles (insets), along
    $C_1$ (left) and $C_2$ (right).}
  \label{fig:mu}
\end{figure}

An easy case which allows to understand the large $S$ behavior of the
HAF on the Kagome lattice consists in a loop of four triangles. In this situation
the classical minimal set is (once accounted for the global $SO(3)$
action) the union of three circles $C_1,C_2,C_3$, two of each crossing at exactly
one point. The crossing points correspond to planar
configurations. There is a symmetry exchanging $C_2$ and $C_3$. In
Figure \ref{fig:mu} we plot the value of $\mu$ along $C_1$ and along
$C_2$, with parameter an angle which is $0$ or $\pi$ on the crossings; this confirms the general belief that $\mu$ is minimal on
planar configurations.

\label{sec:husimi-tree}

The Husimi tree, proposed by Dou\c{c}ot and Simon\cite{doucot_semiclassical_1998}, also serves as a toy model for the study of the Kagome
lattice. It is depicted on Figure \ref{fig:huskag}.

The advantage of this model is that the classical minimal set is much
simpler than on the Kagome lattice. Indeed, on the Husimi tree, once
the three vectors on a parent triangle are chosen along a great equilateral
triangle on the sphere, there is one degree of freedom in the choice
of the spins for each child triangle. Thus the minimal set is a
torus of dimension $\sharp(\text{triangles})-1$, parametrised by the
angles between the equilateral triangles at neighbouring sites.

Dou\c{c}ot and Simon \cite{doucot_semiclassical_1998} reported that the classical frequencies are
not constant on the classical minimal set: in particular, in this
situation there is quantum selection (the selected points are presumed
to be coplanar configurations except for the spins at the leaves which
are free), but there is no thermal selection since there are
equivalent for a class of phase space transformations which preserve
the volume.

\subsection{Anisotropic XXZ chain}
\label{sec:anisotropic-xxz-chain}
Let us take up from an example proposed by Dou\c{c}ot and Simon
\cite{doucot_semiclassical_1998} and define
the following Hamiltonian acting on a closed chain of $N$ spins:
\[
  H=J\sum_i S_i\cdot S_{i+1}+\sum_i S^z_i(S_{i+1}^z-S_{i}^z).
\]
The principal term in the classical energy is
\[
  h_0=J\sum_i \mathbf{e}_i\cdot \mathbf{e}_{i+1}+\sum_i
  z_i(z_i-z_{i+1}).
\]

The next-order contribution is
\[
  h_1=2J\sum_i \mathbf{e}_i\cdot \mathbf{e}_{i+1}+\sum_i
  (-2z_iz_{i+1}+5z_i^2-1).
  \]
If $J<0$, the minimum of $h_0$ is reached on ferromagnetic
configurations $\{\mathbf{e}_i=\mathbf{e}\}$, indexed by
$\mathbb{S}^2$.

Near any of these minimal configurations, the linear spin wave theory
is the same, up to a factor $-J+1-z^2$ in the potential. Hence $\mu$ is
minimal as $z=0$.

On ferromagnetic ordered configurations, one has
\[
  h_1=JN+3Nz^2.
\]
Again $h_1$ is smaller when $z=0$. The sum $\mu+h_1$, which is the
criterion for quantum selection, is minimal as $z=0$, hence the ground
state is located on this set.

%%% Local Variables:
%%% mode: latex
%%% TeX-master: "Main"
%%% End:

\section{Conclusion}
\label{sec:conclusion}

\subsection{Quantum versus thermal selection}
\label{sec:quant-vers-therm}

In this paper, we reported evidence that quantum order by disorder
does not have the same rules as thermal order by disorder. In
experimental settings of low-temperature quantum systems, there is
competition between quantum and thermal selection. We present an
analysis of orders of magnitude.

On the system $SrCrGaO$ which is an experimental realization of the
Kagome lattice, the interaction strength $J$ is presumed\cite{harris_possible_1992} to be of
order
\[
  \frac{J}{k_B}\simeq 50K.
\]
In experimental realizations, the spin $S$ cannot be very large so
that the order of magnitude of the contribution $\mu+h_1$ is also of
order $ 10K\times k_B$. This means that, below these
temperatures, quantum selection predominates over thermal selection,
since the magnitude of the quantum fluctuations is much greater. On
the Kagome lattice there is no competition presumed between quantum
and thermal selection, but in other cases it could even lead to a
phase transition from thermal order (at medium temperatures) to
quantum order (at very low temperatures). The temperatures
involved in this analysis can be reached for large systems by modern
experimental methods.

In our recent paper \cite{deleporte_low-energy_2016} we also computed the relative contributions
at low temperature on systems for which the quantum selection
criterion $\mu+h_1$ is minimal at two points, one of which is a
regular ``miniwell'' point, the other a crossing point. In this
situation, if the temperature is such that thermal effects are of the
same order as quantum effects, then the crossing point will be
selected (the quantum fluctuations do not see the difference between
the two points, and the thermal fluctuations select the one with
maximal local zero dimension). However at lower temperatures, the regular point will be
selected. The interpretation is that $\mu+h_1$ acts as an effective
Hamiltonian, which is smooth on the miniwell, but which is
typically non-regular at the crossing point (see Figure \ref{fig:mu}). This
confinement leads to an increased quantum energy (this shift is of
order $S^{-4/3}$). Hence there are more low-energy quantum states near
the miniwell than near the crossing point. This is a theoretical
instance of a phase transition, which is of course very peculiar
(since $\mu+h_1$ reaches the same value at two very different points).

\subsection{Selection on the Kagome lattice}
\label{sec:select-kagome-latt}
The actual computation of $\mu $ on examples, even as simple as a
chain of triangles, requires the full diagonalization of a matrix whose size
grows with the number of spins, at each minimal point. Variational
approaches allow to show that special (usually planar) configurations
are critical points for $\mu $ (the first derivative of $\mu $
vanishes at these points), but to show that these configurations are
global minima requires additional techniques.

As illustrated in Section \ref{sec:toy-models}, the local geometry of
the minimal classical set plays a very important role. Points near
which the classical minimal set is a smooth manifold are now quite
well understood from a mathematical point of view. On a point where
exactly two manifolds cross, there is a chance that quantum order by
disorder selects the crossing, especially in symmetrical situations
for which the function $\mu$ reaches a local minimum at the
crossing. Conversely, if three or more manifolds cross at a point,
with model the boundary of a hypercube, then the crossing point has no
reason to be selected by the quantum system.

We believe that, near planar
configurations on the Kagome lattice, the local structure of the
classical minimal set is a direct product of structures with two
manifolds crossing \footnote{An example of such a direct product is the
  Schrodinger opertor on $\mathbb{R}^4$ with potential
  $x^2y^2+z^2t^2$. At the point $0$ four manifolds
  cross as a cartesian square of the crossing of two manifolds at a
  point, not as the corner of a hypercube.}, with quartic
non-degenerate part (that is, they follow the model case
above). Indeed, the quadratic and quartic terms
in the energy, near a planar configuration, do not depend on the particular planar configuration, so
that as soon as for one configuration one has a product of structures
as above, it is the case for all configurations.

On systems where the classical minimal set is non smooth, such as the
Kagome lattice, the parametrisation of this set is already a
challenge. Numerical techniques which do not involve knowledge of
the minimal set should be of help in tackling this problem.

\subsection{Tunnelling}
\label{sec:tunnelling}

To conclude with, we address the issue of exponential precision in
estimates related to Toeplitz operators. This problem is relevant in
the context of tunnelling: it is generally hoped that, in the
presence of symmetries, the ground state will tunnel between various
configurations, and the spectral gap (or the inverse time needed for a
quantum state to go from one configuration to another) will be of
order $\exp(-cS)$ in the large spin limit, where $c$ is a ``tunnelling
rate'', related to some classical action.

Various
attempts\cite{garg_macroscopic_1992,awschalom_macroscopic_1992,garg_macroscopic_1990,chudnovsky_quantum_1988,auerbach_small-polaron_1991,anderson_ordering_1956,delft_spin_1993} have been made to study this phenomenon in the setting of spin systems, mainly by removing two antipodal
points on the phase space (the sphere), thus formally transforming the
phase space into $\R\times \S^1$ in which usual (Weyl) quantization
takes place with quantum state space $L^2(\S^1)$. However, it is
doubtful that these attempts yield the correct tunnelling rate. First,
this manipulation changes the quantization procedure, and it is
unclear whether there is a way to perform the computations which
is consistent with the initial problem up to an error of order
$S^{-1}$, let alone an exponentially small error. Second, rates of
decay of order $\exp(-c\hbar^{-1})$ are notoriously delicate
even in the simplest geometrical setting of Weyl quantization on
$\R^{2n}$, as detailed by
Martinez\cite{martinez_introduction_2002}. The basic difficulty is
that one needs to extend data in complex space, which can be done only
if the classical energy is real analytic, and only to a small distance
from the real space. This puts a limit on the actual
tunnelling rate.
Lower bounds (Agmon estimates) on the tunnelling rate for Toeplitz
operators were recently obtained by the author\cite{deleporte_toeplitz_2018}.

%%% Local Variables:
%%% mode: latex
%%% TeX-master: "Main"
%%% End:

\appendix*
\section{Computation of Toeplitz operators on the sphere}
For the particular case of the sphere, one can build Toeplitz
operators as on $\mathbb{C}^d$ via the stereographic projection, which
maps the sphere minus the north pole onto $\mathbb{C}$ in a
holomorphic way.

Via this transformation, quantum states are holomorphic functions on
$\mathbb{C}$  which have finite norm under the following Hermitian structure:
\[
  \langle f|g \rangle_{H_k}=\frac{k+1}{\pi}\int_{\mathbb{C}}
  \frac{\overline{f}(w)g(w)}{(1+|w|^2)^{k+2}}dw.
\]
The space $H_k$ consists of polynomials of degree less than $k$. As
for the flat case, the monomials are orthogonal but not normalized. A
Hilbert basis of $H_k$ is given by
\[
  e_{j,k}:z\mapsto \sqrt{\cfrac{k+1}{\pi}\binom{k}{j}}z^j.
\]

To prove this (and perform further computations), we use the fact
that, for $0\leq j\leq k$:
\[\int_0^{+\infty}\cfrac{u^j}{(1+u)^{k+2}}=\cfrac{1}{k+1}\binom{k}{j}^{-1}.\]
Let us compute the Toeplitz quantization of simple functions defined
on the sphere.

The height $z$ is mapped, via the stereographic projection, in the map
\[
  w\mapsto \cfrac{|w|^2-1}{|w|^2+1}.
\]
Since this function is radial, the matrix elements $\langle
e_{j,k}|T_k(z)|e_{j',k}\rangle_{H_k}$ are zero for $j\neq
j'$. Moreover,
\[\langle e_{j,k}|T_k(z)|e_{j,k}\rangle_{H_k}=\cfrac{2j-k}{k+2}.\]
Hence, in this basis, the operator $T_k(z)$ is $\cfrac{k}{k+2}$ times
a diagonal operator with equidistributed diagonal values from $-1$ to
$1$; that is, the spin operator $S_z$ with $2S=k$. The states
$e_{j,k}$ corresponds to spin states $|S,m\rangle$ with $m=j-S$.

The abscissa $x$ is mapped, via the stereographic projection, into the
map
\[
  w\mapsto \cfrac{2Re(w)}{1+|w|^2}.
\]
This is the sum of a function of winding number $-1$ and a function of
winding number $1$. Hence the matrix of $T_k(x)$ in the natural basis
is zero except on the over- and underdiagonal. The matrix elements are
\[\langle
  e_{j,k}|T_k(x)|e_{j+1,k}\rangle_{H_k}=\cfrac{\sqrt{(k-j)(j+1)}}{k+2}.\]

In this basis the matrix of the operator $T_k(x)$ is
$\frac{k}{k+2}S_x$.

By this method, the Toeplitz quantization of any polynomial in the
coordinates can be computed; this yields Table \ref{tab:sphere}.

\bibliographystyle{aipauth4-1}
\bibliography{SelectionRules}

\end{document}